\documentclass{llncs}
\usepackage{xspace}
\usepackage{latexml}
\def\ebook{\mbox{E-book}\xspace}
\def\ebooks{\mbox{E-books}\xspace}
\def\epub{\textsc{epub}\xspace}
\def\svg{\textsc{svg}\xspace}
\def\zip{\textsc{zip}\xspace}
\def\tikz{\texttt{TikZ}\xspace}

\title{\ebooks and Graphics with \LaTeXML}
\author{Deyan Ginev\inst{1} \and Bruce~R.~Miller\inst{2} \and Silviu Oprea\inst{3}}
\institute{Computer Science, Jacobs University Bremen, Germany
 \and National Institute of Standards and Technology, Gaithersburg, MD, USA
 \and Department of Computer Science, University of Oxford, Oxford, UK}
\date{\today}

\begin{document}
\maketitle
\begin{abstract} 
Marked by the highlights of native generation of \epub \ebooks and \tikz support for creating \svg images, we present an annual report of {\LaTeXML} development in 2013. {\LaTeXML} provides a reimplementation of the {\TeX} parser, geared towards preserving macro semantics;
it supports an array of output formats, notably \HTML5, \epub, {\XHTML} and its own \LaTeX-near \XML.
Other highlights include enhancing performance when used inside high-throughput build-systems, via incorporating a native \zip archive workflow, as well as a simplified installation procedure that now allows to deploy LaTeXML as a cloud service. To this end, we also introduce an official plugin-based scheme for publishing new features that go beyond the core scope of LaTeXML, such as web services or unconventional post-processors.
The software suite has now migrated to GitHub and we welcome forks and patches from the wider FLOSS community.
\end{abstract}

\section{Introduction}
Another busy year of {\LaTeXML}\footnote{see \url{http://dlmf.nist.gov/LaTeXML/}} development has gone by;
while we've not completely accomplished all the tasks we'd hoped for (c.f. \cite{GinMil:latexmlCICM13}),
we've finished others including some we hadn't originally planned.
While it was originally developed for NIST's Digital Library of Mathematical Functions\footnote{see \url{http://dlmf.nist.gov}},
where it continues to serve, we continue to find additional applications.
One, carried out this year, was the natural extension of the system to generate \epub documents;
the first converter, to our knowledge, natively generating \epub\ from \TeX.  
Including \MathML, along with Daisy\footnote{see \url{http://www.daisy.org/}}
support of audio rendering of math, \epub is a major step forward for
accessibility. 
Two planned milestones were also completed, namely: supporting the \tikz, a large, elaborate
graphics package in which one draws complex diagrams, plots and other 2D and 3D graphics
using \TeX\ markup; as well as completing a community-facing project reorganization.
Together, these features are hoped to extend the reach of MKM technologies.

Before we delve into details, a little background about \LaTeXML\ may be in order.
Two main approaches are currently used to generate
\HTML\ from \TeX. The first, exemplified by \texttt{tex4ht}, uses the actual \TeX\ engine
to process the source while redefining certain commands to drop
\verb|\special| data into the normal \texttt{dvi} output.
An alternative \texttt{dvips} then deciphers that augmented \texttt{dvi}
to infer and construct the appropriate \HTML.
In the second approach, used by \LaTeXML, the program
emulates \TeX\ for the most part but interprets some macros
specially, producing \XML\ directly.

The first approach has the advantage of (usually)
allowing the processing of arbitrary \TeX\ and \LaTeX\ packages,
although the resulting \HTML\ may not reflect the intended
structure nor semantics.
The challenges are in the \TeX\ programming necessary to
insert the \verb|\specials|, generating valid, indeed even well-formed, \HTML,
and in recovering sufficient semantic structure from the \texttt{dvi}.

The second approach gives more direct control of the generated output.
It is easier to extend to new \XML\ structures,
and being fundamentally {\XML} aware, it produces valid \XML.
\LaTeXML\ uses an intermediate \XML\ format preserving the semantic structure. 
A feature of \LaTeXML\ `bindings' (\LaTeXML's re-implementation
of \LaTeX\ packages) is control sequences defined to be ``Constructors'',
directly constructing the \XML\ representation of their content.
The challenge lies in emulating \TeX\ sufficiently well to
process complex packages, or alternatively, to
develop \LaTeXML-specific bindings for them.

In either approach, \LaTeX\ packages that define
macros with semantic intent must be dealt with
individually or else the semantics will be lost.

\section{Reorganization}\label{reorganization}
We have reorganized both our code development and our code base.
In the first sense, we have moved our repository to GitHub\footnote{see \url{https://github.com/brucemiller/LaTeXML}} 
where you can more conveniently browse our code, or obtain the latest version.
We have also ported our Trac tickets to GitHub's Issues,
preserving all bug and feature requests.

Along with the move to GitHub came opportunities to share
code and development calling for clearer coding standards.
We committed to code quality and formatting by
adopting \texttt{perltidy} and \texttt{perlcritic} policies,
which were adapted to the polyglot of \TeX, Perl, \XML,
\XSLT, automatically enforced by \texttt{git}.

In the second sense, we have reorganized the code itself to more clearly
separate the modules related to the separate phases of processing.
At the same time, we enable ``conversion as an API'', offering a connection and code sharing between those phases when more
complex processing is called for, such as carrying a single \TeX\ source
file through the full processing to \HTML, or even \epub (see \S\ref{ebooks}).
In particular, it provides better support for daemonized processing, foundational to batch conversions and web service deployments.

This reorganization positions us to develop a plugin architecture allowing modular extensions
covering both new \LaTeX\ styles and bindings, but also
enhanced postprocessing for more sophisticated applications such as s\TeX.
We have already refactored three flavors of {\LaTeXML} web servers,
an alternative grammar for math parsing,
as well as an extension for converting {\TeX} formulas into queries for the MathWebSearch search engine, all hosted on GitHub as separate repositories.
The true power of the new contribution model is revealed when combined with Perl's CPAN distribution and dependency management system,
 which will allow for single command installation of any LaTeXML-based project and its full dependency tree.

\section{\ebooks}\label{ebooks}
The newest version of \epub, version 3, is primarily a packaging
of \HTML\ pages representing chapters or sections into a structured
\zip archive. The big step forward for the scientific community
is that it now calls for the use of \MathML\
to represent mathematics. Since \LaTeXML\ is already generating \HTML,
with embedded \MathML, and allows that output to be split into
multiple pages as specified by the user, it seemed an obvious
and natural extension to generate \epub documents. Moreover, the
web-service spin-off projects had already called for and drafted the compression of the resulting
directory of generated content into a \zip
archive.  Thus, with appropriate rearrangement of the pieces,
and the addition of a Manifest of the correct structure,
we have all the basic components needed to generate \epub documents. We have generated a number of \epub documents and successfully validated them against the official \texttt{idpf} validator\footnote{see \url{http://validator.idpf.org/}}.

We subsequently considered to also add support for Amazon's proprietary \texttt{mobi} \ebook format. However, at the time of writing the \texttt{mobi} ecosystem is transitioning to the new Amazon Kindle Format 8 (\texttt{AKF8}), which aims to more fully align with \epub 3.0. Finally, the lack of an open ecosystem around the format prevented us from repeating the quick and painless design process for the \epub output, so we did not venture further.


\section{Graphics}\label{graphics}
Given the challenges of developing \LaTeXML\ bindings for complex
\LaTeX\ packages, we were skeptical when Michael Kohlhase
initially posed the challenge: Was \LaTeXML's engine good
enough to implement the \tikz package and generate \svg?
The package is so large and complex, not to mention
its development so fast-moving, that creating \LaTeXML-specific
bindings for all its many commands is impractical.  However,
\tikz\ is designed to pass all processed graphics through
a relatively small driver layer, and in fact already has
a \texttt{tex4ht} driver for producing \svg!
Providing we can faithfully emulate all the \TeX\ processing
that leads to that driver layer, we may have a chance;
presumably any semantics implied by \tikz markup isn't so critical,
but the expected \svg obviously is.

The main tasks, then, were to implement \LaTeXML\ bindings
for just that driver, covering universal graphics primitives such as points, lines and angles;
then improve \LaTeXML's engine to cope
with the sophisticated \TeX\ macro usage in the higher
layers of \texttt{pgf} and \tikz.

Ultimately, we succeeded beyond our expectations.
Although the results are not perfect,
\LaTeXML\ now successfully processes 3/4 of the
first page\footnote{see \url{http://www.texample.net/tikz/examples/all/}} of \tikz examples on the
{\TeX}ample.net website, generating valid
\HTML5, with text and \MathML\ combined.
In contrast, \texttt{tex4ht} succeeds on slightly more
than half the examples, often producing invalid markup,
and doesn't support \MathML\ embedded in the \svg.
It must be admitted, however, that \LaTeXML\ is \emph{very}
slow at processing \tikz markup!
Converting the `signpost' example from \TeX ample.net
required almost 2 minutes, whereas \texttt{tex4ht} needed
only 2 seconds (albeit with incomplete math).
\texttt{pdflatex} converts it to pdf in less than half a second.

In the process, we have further improved the
fidelity of the \TeX\ emulation, introduced
a (currently very rudimentary) mechanism for estimating
the size of displayed objects and exercised the
integration of both \MathML\ and \svg\ into \HTML. 
Additionally, {\LaTeXML} now has its own {\TeX} profiler, which
offers binding developers per-macro feedback on
exclusive runtimes, helping to identify core conversion bottlenecks.
These improvements are beneficial even outside the graphics milestone and
contribute to an overall better {\LaTeXML} ecosystem.

Areas needing further work are \tikz' matrix
structure which currently clashes with \LaTeXML's handling
of alignments; inaccuracies of \LaTeXML's sizing of objects;
and, of course, examples involving other exotic packages
not yet known to \LaTeXML.  We plan to test against
the entire suite of examples at {\TeX}ample.net to discover
other weaknesses and further improve the module.

Beyond \tikz, we are hoping to leverage this experience and apply
it to supporting the \texttt{xy} package, another
popular and powerful system.  It seems to have a less
well-defined driver layer and we are in the early stages of
discovering the smallest set of macros that could serve that
function. Nevertheless, we have had some preliminary, proof-of-concept, success.
We already have minimal support for the \texttt{pstricks}
package, but with its Postscript oriented design,
it is more time consuming to develop further bindings.

\section{Outlook}
The initial success with \tikz processing is quite
gratifying, but it needs refinement, and we look forward to testing
on a larger scale. We also intend
to extend our reach to the \texttt{xy} packages.
Other {\ebook} formats such as \texttt{AKF8} should be possible with
specializations of manifest generation and other fine tuning.
Surprisingly, generating Word and OpenOffice formats shares many features
with {\ebook}s; of course finding the documentation and writing the {\XSLT}
transformations from \LaTeXML's native {\XML} to Word's will be challenging.

Our move to GitHub, the code reorganization and the plugin contribution model should make it easier
for users to use and adapt the system, as well as to contribute
back patches and improvements that will help our development.

\bibliography{kbib/kwarc}
\end{document}